\documentclass[12pt,a4paper]{article}
\usepackage[dvips]{epsfig}
\usepackage{epsf,amsfonts,amssymb}

\newcommand{\eqn}[1]{(\ref{#1})}
\newcommand{\be}{\begin{equation}}
\newcommand{\ee}{\end{equation}}
\newcommand{\ben}{\begin{displaymath}}
\newcommand{\een}{\end{displaymath}}
\newcommand{\bea}{\begin{eqnarray}}
\newcommand{\eea}{\end{eqnarray}}
\newcommand{\bean}{\begin{eqnarray*}}
\newcommand{\eean}{\end{eqnarray*}}

\newcommand{\ba}{\begin{array}}
\newcommand{\ea}{\end{array}}
\newcommand{\bi}{\begin{itemize}}
\newcommand{\ei}{\end{itemize}}

\newcommand{\T}{\Theta}
\renewcommand{\O}{\Omega}

\renewcommand{\t}{\theta}

\newcommand{\caln}{\mbox{${\cal N}$}}

\newcommand{\calr}{\mbox{${\cal R}$}}

\newcommand{\bfx}{\mbox{\boldmath $x$}}

\newcommand{\bbe}[1]{{\mathbb E}^{#1}}
\newcommand{\bbr}[1]{{\mathbb R}^{#1}}

\newcommand{\ads}[1]{$\mbox{AdS}_{#1}$}
\newcommand{\adss}[2]{$\mbox{AdS}_{#1}\times {S}^{#2}$}
\newcommand{\rn}{Reissner-Nordstr\"{o}m }

\newcommand{\ket}[1]{\mbox{$| #1 \rangle$}}

\newcommand{\sect}[1]{\setcounter{equation}{0}\section{#1}}

\newcommand{\atmp}[3]{{\it Adv. Theor. Math. Phys.} {\bf #1} {(#2)} #3}
\newcommand{\cmp}[3]{{\it Commun. Math. Phys.} {\bf #1} {(#2)} #3}
\newcommand{\ijmpa}[3]{{\it Int. J. Mod. Phys.} {\bf A #1} {(#2)} #3}
\newcommand{\jhep}[3]{{\it J. High Energy Phys.} {\bf #1} {(#2)} #3}
\newcommand{\mpla}[3]{{\it Mod. Phys. Lett.} {\bf A #1} {(#2)} #3}
\newcommand{\npb}[3]{{\it Nucl. Phys.} {\bf B #1} {(#2)} #3}
\newcommand{\plb}[3]{{\it Phys. Lett.}{\bf B #1} {(#2)} #3}
\newcommand{\prd}[3]{{\it Phys. Rev.} {\bf D #1} {(#2)} #3}
\newcommand{\prep}[3]{{\it Phys. Rep.} {\bf #1} {(#2)} #3}
\newcommand{\prl}[3]{{\it Phys. Rev. Lett.} {\bf #1} {(#2)} #3}

\newcommand{\hepth}[1]{{\tt hep-th/#1}}

\begin{document}

\begin{titlepage}

\bigskip
\rightline{}
\rightline{DAMTP-2001-52}
\rightline{\hepth{0106008}}
\bigskip\bigskip\bigskip\bigskip
\centerline{\Large \bf {Brane Splitting via Quantum Tunneling }} 
\bigskip\bigskip
\bigskip\bigskip

\centerline{\large Selena Ng${}^*$ and Malcolm Perry${}^{\dagger}$}
\bigskip

\centerline{\em Department of Applied Mathematics and Theoretical Physics,} 
\centerline{\em Centre for Mathematical Sciences,}
\centerline{\em Wilberforce Road, Cambridge CB3 0WA, United Kingdom}
\medskip
\bigskip\bigskip

\begin{abstract}
We study the two-centred \adss{7}{4} solution of eleven-dimensional
supergravity using the Euclidean path-integral approach, and find that
it can be interpreted as an instanton, signalling the splitting of the
throat of the M5 brane.  The instanton is interpreted as indicating a
coherent superposition of the quantum states corresponding to
classically distinct solutions.  This is a surprising result since it
leads, through the AdS/CFT correspondence, to contradictory
implications for the dual $(2,0)$ superconformal field theory on the
M5 brane. 
We also argue that similar
instantons should exist for other branes in ten- and
eleven-dimensional supergravity. The counterterm subtraction technique
for gravitational instantons, which arose from the AdS/CFT
correspondence, is examined in terms of its applicability to our
results. Connections are also made to the work of Maldacena et al on
anti-de Sitter fragmentation.  
\end{abstract}

\vfill

\footnoterule
{\footnotesize ${}^*$email: S.K.L.Ng@damtp.cam.ac.uk \vskip -12pt} \vskip 10pt
   {\footnotesize ${}^{\dagger}$email: M.J.Perry@damtp.cam.ac.uk \vskip -12pt}

\end{titlepage}

%%%%%%%%%%%%%%%%%%%%%%%%%%%%%%%%%%%%%%%%%%%%%%%%%%%%%%%%%%%%%%%%%%%%%%%%%%%%%%
%%%%%%%%%%%%%%%%%%%%%%%%%%%%%%%%%%%%%%%%%%%%%%%%%%%%%%%%%%%%%%%%%%%%%%%%%%%%%%
%%%%%%%%%%%%%%%%%%%%%% SECTION 1 %%%%%%%%%%%%%%%%%%%%%%%%%%%%%%%%%%%%%%%%%%%%%
%%%%%%%%%%%%%%%%%%%%%%%%%%%%%%%%%%%%%%%%%%%%%%%%%%%%%%%%%%%%%%%%%%%%%%%%%%%%%%
%%%%%%%%%%%%%%%%%%%%%%%%%%%%%%%%%%%%%%%%%%%%%%%%%%%%%%%%%%%%%%%%%%%%%%%%%%%%%%
%\newpage
\sect{Introduction}
\label{intro}

%%%%%%%%%%%%%%%%%%%%%%%%%%%%%%%%%%%%%%%%%%%%%%%%%%%%%%%%%%%%%%%%%%%%%%%%%%%%%%

Several years ago, Brill \cite{Bri92} considered an instanton
connecting \adss{2}{2} to
a geometry containing two or more \adss{2}{2} centres.  Since \adss{2}{2} is the near-horizon geometry of
the extremal \rn black hole, Brill argued that the
instanton describes the semi-classical splitting of the \adss{2}{2}
throat into two or more throats.  It is well known that the extremal
\rn black hole, and independently its throat \adss{2}{2}, can be
considered as supersymmetric solutions of four-dimensional
supergravity.  In ten and eleven dimensions there exist analogous
supersymmetric solutions of the supergravity action with anti-de Sitter near-horizon geometry, such as the D3 (\adss{5}{5}), M2 (\adss{4}{7}) and M5 (\adss{7}{4})
branes.  The natural question which then arises is
whether the fragmentation process also occurs here in the context of
string or M theory.

One might argue that fragmentation is forbidden on the grounds that
each of these \adss{p+2}{D-p-2} spacetimes is a supersymmetric solution of
supergravity in $D$ dimensions, and thus
stable.  However, rather than indicating an instability,
the existence of such an instanton indicates a quantum superposition
of states.   One expects
that each classical solution of string or M theory is an approximation
to the corresponding quantum state.  Here we will consider
the family of classical solutions with $n$ \adss{p+2}{D-p-2} centres\footnote{We will henceforth refer to these as $n$-centred
\adss{p+2}{D-p-2}, although note that the geometry for $n\neq 1$ is not strictly a
direct product, only approaching one-centred \adss{p+2}{D-p-2}
asymptotically.} (for given $p$ and
$D$), which can be thought of as analogous to
the classical vacua of Yang-Mills theory labelled by winding numbers
$n$. In four-dimensional Yang-Mills theory, the existence of
instantons connecting two vacua with two given winding numbers is 
interpreted as meaning that the true quantum vacuum, often referred to
as the theta vacuum, is a linear superposition of the quantum states
associated to each of the classical vacua.  By analogy, we should
interpret the instanton as indicating that the quantum state
$\ket{\psi_n}$ associated to the $n$-centred
\adss{p+2}{D-p-2} geometry is not necessarily an exact eigenstate of the
M theory ``Hamiltonian''.  Instead we expect an exact eigenstate
to be a coherent superposition of all the states
$\{\ket{\psi_n}\}$.  

However, a second and more serious objection to higher-dimensional AdS
fragmentation is due to the AdS/CFT correspondence \cite{Mal97, GKP98,
Wit98a}.  According to this duality, \ads{2} is dual to quantum
mechanics, and fields in quantum mechanics do not have well-defined
vacuum expectation values, that is, they can fluctuate.  Hence one
would expect that the wavefunction describing the quantum state should be
constructed as a superposition of each of the different classical
vacua, which is in agreement with the existence of Brill's instanton.
On the other hand, higher-dimensional AdS spaces are dual to quantum
field theories.  Scalar fields in field theories can, in contrast to
quantum mechanics, be given fixed vacuum expectation values which are
classical and constant.  Since it costs an infinite action in field
theory to change such vacuum expectation values dynamically,
superposition of different exact vacua in the moduli space of the
theory is forbidden.  Translating to the AdS side, this would seem to
imply that higher-dimensional analogues of the Brill instanton should
not be allowed, if the AdS/CFT correspondence holds true.

Having established the above objections, it is somewhat surprising
that we do identify in this paper, analogously to the four-dimensional case, an
instanton which interpolates between the near-horizon \adss{7}{4}
geometry of the M5 brane and the ``near-horizon''\footnote{We mean this
in the sense of \cite{MMS98} where the limit $L_p\rightarrow 0$ is
taken, and $L_p$ is the Planck length ($L_p^{D-2} = G_D$).}
two-centred \adss{7}{4} geometry of the two-centred M5 brane.  We
postpone discussion on the resolution of this puzzle to the final section.  

The standard procedure for identifying instantons, which we follow, is
to use the Euclidean path
integral approach to quantum gravity \cite{Haw79}.  We will be interested in
contributions to the zero-temperature vacuum amplitude, given by the
path integral
\be
Z=\int D[\phi] e^{-I[\phi]}
\ee
over all fields $\phi$ which are real
on the Euclidean section and with boundary conditions appropriate to
the zero-temperature vacuum, and where $I$ is the Euclidean action. 
In the semi-classical approximation, these contributions correspond to
{\it gravitational instantons}, that is, nonsingular and geodesically
complete solutions of the Euclidean equations of motion with finite action.

The Euclidean action is defined, for a metric $g_{ab}$ and $(p+1)$-form
gauge potential $A_{p+1}$ on a $D$-dimensional manifold $ M$ with
boundary, as 
\be
I = -\frac{1}{\kappa_D^2} \int_{ M} d^Dx \sqrt{g}\left(R -
\frac{1}{2\cdot(p+2)!} F_{\it p+2}^2\right) - \frac{2}{\kappa_D^2}
\int_{\partial M} d^{D-1}x K\sqrt{h} - I_0[h].
\label{gen-action}
\ee
$R$ is the Ricci scalar of the metric, $F_{\it p+2}=dA_{\it p+1}$ is the
$(p+2)$-form field strength, $h_{ab}$ is the induced metric on the
boundary, $K$ is the extrinsic curvature, and $\kappa_D^2~=~16\pi G_D$
where $G_D$ is the $D$-dimensional gravitational constant.  The
surface integral is necessary in order to obtain an action which
depends only on first derivatives of the metric and not normal
derivatives to the boundary \cite{Yor72, GH77},
and ensures that the variational principle is well-defined.  Finally,
$I_0[h]$ is another surface term which depends only on 
the induced metric $h_{ab}$, and whose purpose will become clearer below.

Since the volume of anti-de Sitter space is infinite, the action
\eqn{gen-action} evaluated on an asymptotically \adss{p+2}{D-p-2} space
generally diverges. A common
solution to this problem is to first regularise the action by imposing
cut-offs, and then
subtract the action $I_0$ of a background which acts as a
``zero-point'' and contains the same infrared divergences \cite{HP83,
Wit98b}.  This zero-point action must be written as a surface integral
and corresponds exactly to the term $I_0[h]$ in \eqn{gen-action}
above.  However this background subtraction technique is not in
general well-defined.  If for a particular geometry there exists a
reference background with the same intrinsic metric on the boundary,
the process works.
Unfortunately, one cannot always find such a reference background.  This is
indeed the case for the multi-centred \adss{p+2}{D-p-2} spacetimes. 

The counterterm approach \cite{HS98,BK99, EJM99}, which
was motivated by the AdS/CFT correspondence, provides an alternative and well-defined procedure.  Since
the partition function of the string theory is conjectured to be equal
to the generating functional of the dual conformal field theory, one
can remove the divergences in the supergravity action  by adding local
counterterms on the boundary, giving a manifestly-finite, renormalised
on-shell action.  These local counterterms are proportional to surface
integrals of the induced metric, and hence also take the form of the
term $I_0[h]$ in \eqn{gen-action}. As a side-remark, a consequence of
this is that any interpolating, nonsingular, asymptotically AdS
solution of the Euclidean field equations will have finite action, and
can be interpreted as a gravitational instanton. 

Much of the literature on using counterterm subtraction in Euclidean
quantum gravity has focused on gauged supergravities obtained by reduction on a
sphere of the ten- or eleven-dimensional
theory.  However, the multi-centred \adss{p+2}{D-p-2} solutions are
{\it warped} geometries, due to the choice of direction connecting the
two centres.   This means that, as far as we are aware, the
$D$-dimensional theory cannot be consistently reduced to a
$(p+2)$-dimensional gauged supergravity, and thus the existing
counterterms in \cite{Ske00} for asymptotically AdS spaces cannot be
faithfully applied.  An alternative would be to attempt to formulate the
counterterms directly in the higher dimensions, and recent progress in
this area has been made in \cite{Tay01, Tay01a}.  This will be discussed further in section~\ref{counterterms}.

In the next section we will review the work of Brill on \adss{2}{2},
and in section~\ref{M5-brane} we present the \adss{7}{4} instanton for the M5 brane using background subtraction.  We will also compare
our results with the claim in a paper by Maldacena et al \cite{MMS98}
that brane creation by the field strength in supersymmetric AdS spaces
occurs only for \ads{2}.  In section~\ref{counterterms} we examine the
limitations of the background subtraction procedure, and discuss how
these might be overcome by suitable counterterm
regularisation. Section~\ref{other} contains preliminary results for
the M2 and D3 branes, and a more conclusive result for the D1/D5
system, including a short discussion on correctly accounting for the
contribution from self-dual field strengths to the instanton action.   
We also briefly speculate on nonconformal D branes and the NS5 brane.
In the concluding section, we comment that our result for the M5 brane implies the
superposition of
distinct quantum vacua in the dual six-dimensional
$\caln$=(2,0) superconformal field theory, in contradiction to the
cluster decomposition principle, and speculate on its resolution.  
Note that all quantities, unless otherwise stated, will be presented
in Euclidean signature.

%%%%%%%%%%%%%%%%%%%%%%%%%%%%%%%%%%%%%%%%%%%%%%%%%%%%%%%%%%%%%%%%%%%%%%%%%%%%%%
%%%%%%%%%%%%%%%%%%%%%%%%%%%%%%%%%%%%%%%%%%%%%%%%%%%%%%%%%%%%%%%%%%%%%%%%%%%%%%
%%%%%%%%%%%%%%%%%%%%%% SECTION 2 %%%%%%%%%%%%%%%%%%%%%%%%%%%%%%%%%%%%%%%%%%%%%
%%%%%%%%%%%%%%%%%%%%%%%%%%%%%%%%%%%%%%%%%%%%%%%%%%%%%%%%%%%%%%%%%%%%%%%%%%%%%%
%%%%%%%%%%%%%%%%%%%%%%%%%%%%%%%%%%%%%%%%%%%%%%%%%%%%%%%%%%%%%%%%%%%%%%%%%%%%%%
%\newpage
\sect{The Brill Instanton}
\label{RN-BH}

%%%%%%%%%%%%%%%%%%%%%%%%%%%%%%%%%%%%%%%%%%%%%%%%%%%%%%%%%%%%%%%%%%%%%%%%%%%%%%

For a one-form potential and metric in four dimensions, the
supergravity action \eqn{gen-action} reduces to
\be
I = -\frac{1}{\kappa_4^2} \int_{ M} d^4x\sqrt{g} \left(R-\frac{1}{2\cdot 2!}
F_{\it 2}^2\right) - \frac{2}{\kappa_4^2}\int_{\partial M} d^3 x\, K\sqrt{h} -
I_0[h].
\label{4d-action}
\ee
One family of solutions to the equations of motion of this action is
the metric and two-form field strength
\bea
ds^2 &=& H^{-2}dt^2 + H^2 ds^2(\bbe{3}),
\label{4d-metric} \\
F_{\it 2} &=& 2\, \star_{3} dH , 
\label{4d-form}
\eea
where $H$ is a harmonic function on $\bbe{3}$ and `$\star_{3}$' is the
Hodge dual also on $\bbe{3}$.   One recovers (one-centred) \adss{2}{2}
from the metric \eqn{4d-metric} by choosing
\be
H = \frac{b}{|\bfx|},
\ee
where $b$ is a constant and $\bfx$ represents Cartesian coordinates on
$\bbe{3}$.  The apparent singularity at $\bfx=0$ is a
nonsingular horizon of the metric \eqn{4d-metric}.  In fact, the
metric represents the near-horizon limit of the extremal \rn black hole
with charge and mass $b$.
Since the gravitational attraction between extremally charged black
holes is balanced by the electric repulsion, another
choice is the multi-centred harmonic function
\be
H = \sum_{k=1}^n \frac{b_k}{|\bfx - \bfx_k |} .
\ee
This corresponds to $n$ black holes each of charge (of equal sign)
$b_k$ centred at $\bfx=\bfx_k$.  Geometrically one has a spacetime which tends
asymptotically to \adss{2}{2} with radius $b_k$ whenever one of these centres is
approached, that is, as $\bfx\rightarrow \bfx_k$, and to \adss{2}{2}
with radius $b_1+\ldots + b_n$ as the outer boundary $|\bfx|\rightarrow\infty$ is approached.  This
$n$-centred \adss{2}{2} spacetime thus interpolates between the $n+1$
\adss{2}{2} spaces.  In the following we will restrict ourselves to
the two-centred case $n=2$, however, it is straightforward to extend
the analysis to general values of $n$.

Let us now, following Brill \cite{Bri92}, compute the Euclidean action
\eqn{4d-action} for the two-centred \adss{2}{2} geometry.  As
mentioned in the introduction, one expects the result to be infrared-divergent since \ads{} is noncompact.  We will thus background subtract the reference action of one-centred
\adss{2}{2} in order to obtain a finite, meaningful result.

The metric equation of motion demands that the Ricci scalar
vanishes in the action \eqn{4d-action}, reducing the bulk term to an integral of the two-form
field strength.  The source-free equation of motion $d*F_{\it 2}=0$ for the field
strength implies that one can write $\tilde{F}_{\it 2}\equiv *F_{\it 2} = d\tilde{A}$, for a one-form potential $\tilde{A}$.   So the bulk term can be converted into 
\be
\int_{ M} d^{4}x \sqrt{g}\, \frac{1}{2!}F_{\it 2}^2 =
\int_{ M} F_{\it 2}\wedge *F_{\it 2} 
= \int_{ M} F_{\it 2}\wedge d\tilde{A} 
= \int_{\partial M} F_{\it 2}\wedge \tilde{A} \, + \, C ,
\ee
since $dF_{\it 2}=0$ by the Bianchi identity. Note that although $\tilde{A}$
is only specified up to a gauge transformation, the surface integral
above is obviously gauge-invariant. Here $C$ represents finite
contributions to the action resulting from the non-smoothness of $\tilde{A}$.
Since we are only interested in the finiteness of the action rather
than its specific value, we will not explicitly evaluate these
contributions here.

In spherical coordinates on $\bbe{3}$ such that
\be
ds^2 (\bbe{3}) = dr^2 + r^2(d\t^2 + \sin^2\t\, d\phi^2)
\ee
the two-centred harmonic function takes the form
\be
H = \frac{b_1}{r_1} + \frac{b_2}{r_2} ,
\label{4d-two-H}
\ee
where
\be
r_1^2 = r^2+a^2+2ar\cos\t , \hspace{1cm}  r_2^2 = r^2+a^2-2ar\cos\t .
\label{def-r1r2}
\ee
The two-form field strength, its dual and associated one-form
potential are given by
\bean
F_2 &=& 2 \partial_r H\, r^2  \sin\t \, d\t\wedge d\phi +
2\partial_{\t} H\,\sin\t\, d\phi \wedge dr ,
\label{4d-twoform} \\
\tilde{F}_2 &=& 2H^{-2}\partial_r H dt\wedge dr + 2H^{-2}
\partial_{\t}H dt\wedge d\t, \\
\tilde{A} &=& 2H^{-1} dt .
\eean
Since the boundary of \ads{} lies at $r=\infty$ in these coordinates, we need to regularise the action by
introducing an upper cutoff $R$ on the radial integration,
and an upper cutoff $V_1$ on the temporal integration, then taking the limit
$R\rightarrow\infty$ followed by $V_1\rightarrow\infty$ after performing background
subtraction.  The unit outward normal to the regularised boundary
surface $r=R$ is $n=H^{-1}\partial_r$, giving the extrinsic curvature
term
\be
K\sqrt{h} \equiv n\sqrt{h} = (r^2 H^{-1}\partial_r H + 2r)\sin\t .
\ee
Thus the regularised action can be written in these coordinates as
\bea 
\kappa_4^2 I(R)  &=& \frac{1}{2}\int_{\partial M} F_{\it 2}\wedge \tilde{A} -
2\int_{\partial M} d^3 x\, K\sqrt{h}  - \kappa_4^2 I_0[h] + \frac{C}{2}\\
&=& -4\int_{r=R} d^3x \left(r + r^2H^{-1}\partial_r
H\right) \sin\t + \frac{C}{2}. \nonumber
\eea
Note that this vanishes identically for the one-centred geometry, making the background subtraction trivial.
Substituting in the value of the two-centred harmonic function
\eqn{4d-two-H} and expanding in powers of $1/R$ about $1/R=0$, we obtain the result
\ben
\kappa_4 I(R) = -16\pi V_1 \left[R-(R+O(R^{-1}))\right] + \frac{C}{2},
\een
which is finite in the limit $R\rightarrow \infty$!

There are several points to note here.  Firstly,  although we are only
interested in whether or not the action is finite and not in its exact
value, we should point out that, as noted by Brill \cite{Bri92}, the finite contributions  result in the action giving the expected Bekenstein-Hawking entropy $S=A/4$, where $A$ is the nonzero horizon area of the \rn black hole.  The contribution to the path integral is then roughly $\exp(-2I/\hbar)$ which is infinitely suppressed, as expected, in the classical limit $\hbar\rightarrow 0$.
  
Secondly, note the particular interpretation of the Brill instanton.  In Lorentzian signature, the solution is static in time and inhomogeneous in space.  In the Euclidean continuation, however, one can interpret the solution as being static in space and inhomogeneous in time (with a suitable redefinition of the coordinates as in \cite{Bri92}), since each spacetime direction now carries the same sign.  
 The instanton thus describes being in an initial
state (the one-centred universe) in one  asymptotic region, and
rolling to the final state (the two-centred universe) asymptotically.
 Both the initial and final states, being supersymmetric, are degenerate
minima of the action of the theory --- differing from the so-called ``bounce''
instanton solutions in which the solutions are saddlepoints and not
genuine minima --- and thus the quantum system is expected to be a
superposition of these two states.  

Thirdly, we have taken the {\it magnetic} solution \eqn{4d-twoform}
for the field strength $F_{\it 2}$.  We could also have chosen its electric
dual $\tilde{F}_{\it 2}$.  This leads to a real instanton since in
addition to the Wick rotation $t_L\rightarrow t_E=it_L$, one
analytically continues the electric charge $Q_L\rightarrow Q_E=iQ_L$,
thus ensuring that fields which are real in Lorentzian space correspond to real
fields on the Euclidean section \cite{Haw79}.

Finally, note that the finite action is independent of the
distribution of the charges, since the action vanishes independently of
the choice of harmonic function --- this agrees, by taking the limit
$|b_1| \ll |b_2|$, with the result of
Maldacena et al \cite{MMS98} where a one-centred \adss{2}{2} universe
was shown to fragment into a macroscopic universe and a microscopic
brane.  This can also be viewed as brane creation by the two-form
field strength $F_{\it 2}$ with charge proportional to $b_1$. 

%%%%%%%%%%%%%%%%%%%%%%%%%%%%%%%%%%%%%%%%%%%%%%%%%%%%%%%%%%%%%%%%%%%%%%%%%%%%%%
%%%%%%%%%%%%%%%%%%%%%%%%%%%%%%%%%%%%%%%%%%%%%%%%%%%%%%%%%%%%%%%%%%%%%%%%%%%%%%
%%%%%%%%%%%%%%%%%%%%%% SECTION 3 %%%%%%%%%%%%%%%%%%%%%%%%%%%%%%%%%%%%%%%%%%%%%
%%%%%%%%%%%%%%%%%%%%%%%%%%%%%%%%%%%%%%%%%%%%%%%%%%%%%%%%%%%%%%%%%%%%%%%%%%%%%%
%%%%%%%%%%%%%%%%%%%%%%%%%%%%%%%%%%%%%%%%%%%%%%%%%%%%%%%%%%%%%%%%%%%%%%%%%%%%%%
%\newpage
\sect{The M5 Throat Instanton}
\label{M5-brane}

%%%%%%%%%%%%%%%%%%%%%%%%%%%%%%%%%%%%%%%%%%%%%%%%%%%%%%%%%%%%%%%%%%%%%%%%%%%%%%

As mentioned in the introduction, one might expect that a similar
instanton exists in the context of ten- or eleven-dimensional
supergravity.  It turns out that the simplest case to consider is the throat of the M5
brane in eleven-dimensional supergravity, since it is similarly
nonsingular and couples magnetically to the three-form potential
$A_{\it 3}$
in the theory.  The relevant bosonic action is
\be
\kappa_{11}^2 I = - \int_{ M} d^{11}x \sqrt{g} \left( R - \frac{1}{2\cdot 4!}
F_{\it 4}^2 \right) + \int_M F_{\it 4}\wedge F_{\it 4}\wedge A_{\it 3} - 2\int_{\partial M}
d^{10}x \,
K\sqrt{h} - \kappa_{11}^2 I_0[h].
\label{11d-action} 
\ee
We are interested in the solution
\bea
ds^2 &=& H^{-1/3} ds^2(\bbe{6}) + H^{2/3} ds^2(\bbe{5}),
\label{11d-metric} \\
F_{\it 4} &=& \star_{5} dH .
\eea
$H$ is now a harmonic function on $\bbe{5}$ and `$\star_5$' is the
Hodge dual on $\bbe{5}$. 
The usual M5 brane solution is obtained by taking $H=1+b^3/|\bfx|^3$
where $\bfx$ are Cartesian coordinates on $\bbe{5}$, $b$ is a
constant, and the apparent singularity at $\bfx=0$ again corresponds
to a horizon.  If one discards the requirement of asymptotic flatness,
then one can go directly to the ``throat'' or near-horizon geometry
\adss{7}{4} by setting
\be
H = \frac{b^3}{|\bfx|^3},
\ee
and similarly for the $n$-centred geometry.  The constant $b^3$ is now
proportional to the charge or the number of M5 branes located at
$\bfx=0$, and determines the radius $l$ of \ads{7} by $l=2b$. 

We would like to evaluate the supergravity action for the two-centred
\adss{7}{4} geometry analogously to the previous section.   Choosing
spherical coordinates on $\bbe{5}$, 
\be
ds^2(\bbe{5}) = dr^2 + r^2(d\t^2 + \sin^2\t\, d\O_3^2),
\ee
where $d\O_3^2$ is the metric on the unit 3-sphere, we can
write the two-centred harmonic function as
\be
H = \frac{b_1^3}{r_1^3} + \frac{b_2^3}{r_2^3},
\ee
with $r_1$, $r_2$ defined as in \eqn{def-r1r2}.
The metric equation of motion sets
\ben
R = \frac{1}{6\cdot 4!}F_{\it 4}^2 .
\een
As in the previous section, the bulk term can be converted into a
surface integral. Noting that 
\bean
K\sqrt{h} &=& \left(\frac{1}{3}r^4 H^{-1}\partial_r H + 4r^3\right)
\sin^3\t \, \T_3, \\
F_{\it 4} &=& \partial_r H\, r^4 \sin^3\t \,d\t\wedge \mbox{vol}(S^3)
+ \partial_{\t} H\, r^2 \sin^3\t\, dr\wedge \mbox{vol}(S^3),\\
\tilde{F}_7 &\equiv& *F_{\it 4} = d\tilde{A}_{\it 6} = H^{-2}\partial_r H \,
dr\wedge \mbox{vol}(\bbe{6}) +
H^{-2}\partial_{\t} H\, d\t\wedge \mbox{vol}(\bbe{6}),\\ 
\tilde{A}_{\it 6} &=&  -H^{-1}\, \mbox{vol}(\bbe{6}),
\eean
where $\int \T_3 = \O_3$, the volume of the unit three-sphere, and that
the Chern-Simons term vanishes for the above solution,
we can rewrite the regularised action as 
\bea
\kappa_{11}^2 I(R) &=& \frac{1}{3}\int_{\partial M}
F_{\it 4}\wedge\tilde{A}_{\it 6} - 2\int_{\partial M}d^{10}x K\sqrt{h} -
\kappa_{11}^2 I_0[h] \\
&=& -\int_{r=R} d^{10}x \left( r^4 H^{-1}\partial_r H +
8r^3\right) \T_4  - \kappa_{11}^2 I_0[h]. \nonumber 
\eea
Introducing further upper cutoffs on the brane worldvolume directions by
compactifying them on a six-torus with volume $V_6$, we
obtain
\ben
\kappa_{11}^2 I(R) = - V_6 \O_4 \left( 5R^3 + \frac{9}{5}\Delta_3^2\,
a^2 R \right) + O(R^{-1}) - \kappa_{11}^2 I_0[h] ,
\een
where 
\be
\Delta_k(b_1,b_2)=\frac{b_1^k-b_2^k}{b_1^k+b_2^k}
\label{def-delta}
\ee
describes the charge distribution of the configuration.
This is clearly
divergent in the limit $R\rightarrow \infty$, as we expect.  In order
to overcome this divergence, we will perform background subtraction,
choosing as our ``zero-point'' the one-centred action
\be
\kappa_{11}^2 I_0(R) = -5 V_6 \O_4 R^3 .
\label{ads-7-4-one}
\ee
The background-subtracted action is then
\ben
\kappa_{11}^2 I(R) = -\frac{9}{5}\Delta_3^2 \, V_6 \O_4 a^2 R + O(R^{-1}),
\een
which vanishes in the limit $R\rightarrow\infty$ followed by $V_6\rightarrow\infty$ only in the case that the charges are equally distributed, that is,
$\Delta_3=0$, and otherwise diverges linearly in $R$ (and $V_6$).  

How should we interpret this result?  Analogously to Brill's instanton, with a suitable
choice of coordinates, one can interpret the instanton as tunneling
between a universe with two equally-charged centres and a universe
with the same total charge, again at zero temperature.
As we shall see below, the asymptotic boundary of two-centred
\adss{7}{4} in the case $b_1^3=b_2^3$ matches exactly to the boundary
of one-centred \adss{7}{4} with charge $b^3=b_1^3+b_2^3$, and thus the
background subtraction is well-defined.  This
is not the case when $b_1^3\neq b_2^3$, where the background subtraction
calculation must be taken with caveats.  This means that although the
above calculation suggests that no instanton exists which interpolates
between a two-centred spacetime with $b_1^3\neq b_2^3$ and the one-centred
geometry, we cannot conclusively assert this without applying
a better-defined procedure such as the use of counterterms.
Nevertheless, note that the result from background subtraction is
consistent with the analysis of \cite{MMS98} in which the special case
$|b_1| \ll |b_2|$ is considered. Here they find that fragmentation of
a supersymmetric \ads{} universe into one macroscopic and one
microscopic part, or the creation of a brane with charge given by
$b_1$, is allowed only for \ads{2}, but not for higher-dimensional AdS spaces. 

Note that, as opposed to the Brill instanton, in the case of the M5
brane the finite contribution vanishes since it is proportional to the horizon area, which vanishes for the M5 brane.  Hence the zero action in this case correctly coincides with the zero entropy associated to the M5 brane.

%%%%%%%%%%%%%%%%%%%%%%%%%%%%%%%%%%%%%%%%%%%%%%%%%%%%%%%%%%%%%%%%%%%%%%%%%%%%%%
%%%%%%%%%%%%%%%%%%%%%%%%%%%%%%%%%%%%%%%%%%%%%%%%%%%%%%%%%%%%%%%%%%%%%%%%%%%%%%
%%%%%%%%%%%%%%%%%%%%%% SECTION 4  %%%%%%%%%%%%%%%%%%%%%%%%%%%%%%%%%%%%%%%%%%%%
%%%%%%%%%%%%%%%%%%%%%%%%%%%%%%%%%%%%%%%%%%%%%%%%%%%%%%%%%%%%%%%%%%%%%%%%%%%%%%
%%%%%%%%%%%%%%%%%%%%%%%%%%%%%%%%%%%%%%%%%%%%%%%%%%%%%%%%%%%%%%%%%%%%%%%%%%%%%%
%\newpage
\sect{The Need for Counterterms}
\label{counterterms}

%%%%%%%%%%%%%%%%%%%%%%%%%%%%%%%%%%%%%%%%%%%%%%%%%%%%%%%%%%%%%%%%%%%%%%%%%%%%%%

Let us examine in more detail how closely the boundary geometry of 
two-centred \adss{p+2}{D-p-2} matches that of the one-centred space,
so that we can determine in which cases background subtraction can be
faithfully applied.  We will analyse this briefly for \adss{2}{2} and
\adss{7}{4}, then outline the counterterm technique and discuss its
applicability to the M5 throat instanton action.  

%%%%%%%%%%%%%%%%%%%%%%%%%%%%%%%%%%%%%%%%%%%%%%%%%%%%%%%%%%%%%%%%%%%%%%%%%%%%%%

\subsection{Boundary Matching in Background Subtraction}

For \adss{2}{2}, the induced metric on the boundary is
\be
ds^2 = H^{-2} dt^2 + H^2 R^2 d\O_2^2,
\ee
in the limit $R\rightarrow\infty$, and where $H$ in the above equation
is understood to be a function of the regulatory constant $R$ rather
than the coordinate $r$.  
The $S^2$ radius for two-centred \adss{2}{2}
\ben
HR = \left(\frac{b_1}{R_1}+\frac{b_2}{R_2} \right)R  
\een
tends exactly to the one-centred $S^2$ radius $b$ at the boundary, provided that charge conservation holds,
that is, $b_1+b_2=b$. Note that $R_1, R_2$ are defined as
$r_1, r_2$ in \eqn{def-r1r2} with $r$ replaced by $R$. However, the scale factor of the worldvolume coordinates
\be
H^{-2} = (b_1+b_2)^{-2} [ R^2 +
2\Delta_1\, aR\cos\t + a^2(1-3(1-\Delta_1^2)\cos^2\t) ] + O(R^{-1}) 
\label{wv-factors-brill}
\ee
does not match asymptotically, where $\Delta_1$ is defined as in
\eqn{def-delta}. The leading order term agrees with the one-centred
value of $R^2/b^2$ (with charge conservation),
but one is left with non-vanishing terms for any value of $\Delta_1$,
that is, regardless of charge distribution!  Despite this
non-matching, however, the results in section \ref{RN-BH} suggest that
background subtraction still reproduces the correct result.   We shall
see in a few paragraphs that this is indeed the case.

Let us now turn to \adss{7}{4}.  Here the boundary metric is
\be
ds^2 = H^{-1/3} ds^2(\bbe{6}) + H^{2/3} R^2 d\O_4^2 
\ee
in the limit $R\rightarrow\infty$.
The $S^4$ radius for two-centred space matches that for the
one-centred space in the boundary limit.\footnote{In fact, one can show that
the $S^{D-p-2}$ radius for the two-centred \adss{p+2}{D-p-2} always
tends to the corresponding one-centred value in the boundary limit.}
The worldvolume scale factor
\be
H^{-1/3} = (b_1^3+b_2^3)^{-1/3} [ R + \Delta_3\, a\cos\t ] + O(R^{-1})
\label{wv-factors-M5}
\ee
again does not in general match asymptotically, except if $\Delta_3=0$,
that is, when $b_1^3=b_2^3$!  For unequal distributions, however, one
is left with a finite term which renders the background subtraction
procedure untrustworthy.  So although the results of section
\ref{M5-brane} imply that no instanton exists for a two-centred
spacetime with $b_1^3\neq b_2^3$, we cannot state this definitively
without applying a better-defined procedure.  Such a procedure is, as
mentioned in the introduction, counterterm subtraction.

%%%%%%%%%%%%%%%%%%%%%%%%%%%%%%%%%%%%%%%%%%%%%%%%%%%%%%%%%%%%%%%%%%%%%%%%%%%%%%

\subsection{Counterterms and the M5 Throat Instanton} 

Counterterm subtraction has been employed 
successfully to calculate gravitational actions and thermodynamics of
black holes such as the Taub-NUT-AdS, Taub-Bolt-AdS \cite{EJM99}, and Kerr-AdS
\cite{AJ99}, for which the appropriate reference background was
either unknown or ambiguous.   In each of these cases, background
subtraction provided a good approximation to the
results obtained using counterterms, in some cases coinciding.  

The idea behind this technique is as follows.  The AdS/CFT
correspondence equates the partition functions of the two theories.
In the low energy limit, one can thus use the supergravity action in
string theory to calculate the generating functional of the conformal
field theory on the boundary.  It turns out that the divergences which
appear in the gravitational action are local integrals on the
boundary \cite{Wit98a}, and thus can be removed by subtracting local surface
counterterms.  The form of these counterterms has been developed in
\cite{HS98,BK99}.  In particular, the counterterms
to the purely gravitational action are (for $0\leq p\leq 5$) \cite{EJM99}
\bea
I_{ct} &=& -\frac{2}{\kappa_{p+2}^2}\int_{\partial M}
d^{p+1}x \, \sqrt{h} \left[ \, \frac{p}{l} + \frac{l\calr}{2(p-1)}
\right. 
\label{ct-action} \\
& & \left. \hspace{4.3cm}
+ \, \frac{l^3}{2(p-3)(p-1)^2}\left( \calr_{ab}\calr^{ab} - \frac{p+1}{4p}
\calr^2 \right)  \right] + \, I_{\log} , \nonumber
\eea
where  $M$ is the asymptotically \ads{p+2} space, $\calr$ is the Ricci
scalar of the induced metric $h$ on the boundary $\partial M$, and
$I_{\log}$  is a logarithmically divergent term corresponding to the
conformal anomaly \cite{HS98} which only contributes in dimensions for
which $p$ is odd.  Notice that no counterterms are necessary for
$p=0$, which corresponds to the action for the Brill instanton.  The
action above should be read such that for 
$p=1$ only the first term is included, for $p=2,3$ only the first
two terms are included, and for $p=4,5$ all three terms are included.
In addition, for odd values of $p$ there is a contribution
from the logarithmic term $I_{\log}$.  Recall that $l$ is the
radius of AdS.  

For the \adss{7}{4} instanton, the relevant counterterm action is
\be
I_{ct}(\mbox{\ads{7}}) =  -\frac{2}{\kappa_7^2}\int_{\partial AdS_7} d^6 x
\sqrt{h} \left[ \frac{5}{2b} + \frac{b\calr}{4} + \frac{b^3}{8}
\left(\calr_{ab}\calr^{ab} - \frac{3}{10}\calr^2\right) \right] + I_{\log} ,
\label{ct-action-M5}
\ee
since $l=2b$ for the \adss{7}{4} metric \eqn{11d-metric}.
As a first check we compute the counterterms for the one-centred
\adss{7}{4} action.  Since the boundary of \ads{7} is
Ricci-flat, we have that $I_{log}$ vanishes \cite{HS98} and
\ben
I_{ct}(\mbox{\ads{7}}) = -\frac{2}{\kappa_7^2}\int_{r=R} d^6x
\sqrt{h} \frac{5}{2b} = -\frac{V_6}{b^4 \kappa_7^2} 5R^3 .
\een
To uplift to eleven dimensions we simply use that $\kappa_{11}^2 =
\kappa_7^2 \,b^4 \O_4$,
giving
\ben
I_{ct}(\mbox{\adss{7}{4}})= -\frac{V_6 \O_4}{\kappa_{11}^2} 5R^3 ,
\een
which is exactly the value of the one-centred \adss{7}{4} action
obtained in \eqn{ads-7-4-one}, as we expect.  

The counterterms presented in \eqn{ct-action-M5} arise from including
a source for the stress-energy tensor in the dual conformal field
theory, that is, they only cancel divergences corresponding to the
purely gravitational action.  However, the supergravity action
\eqn{11d-action} also contains matter coupled to gravity in the form
of the field strength $F_{\it 4}$.  One might wonder if additional
counterterms are necessary.  To check this, one should first determine
which seven-dimensional fields correspond to the eleven-dimensional
$F_{\it 4}$.  This can be done since the Kaluza-Klein truncation of 
eleven-dimensional supergravity to maximally
gauged supergravity in seven dimensions with gauge group $SO(5)$ has
been shown to be consistent in \cite{NVV99}.  One should then apply
the counterterm criterion for the seven-dimensional scalar fields and
other matter, which was derived in 
\cite{DSS00,Tay00} by turning on a source for each dual operator in
the conformal field theory and following the same approach as for pure
gravity.  Without explicitly undertaking this procedure, however, the
fact that the desired matching is obtained by adding the gravitational
counterterm alone suggests that further counterterms should not be required.  

We would like to compute counterterms for the two-centred action.
However this is not straightforward.  The reason is that 
the counterterms presented in 
\cite{HS98, BK99, EJM99} have been formulated for asymptotically AdS
spacetimes.  In the case that the solution of interest takes the form $(\mbox{asymptotically \ads{p+2}})\times
S^{D-p-2}$, one can in general reduce the $D$-dimensional theory to
$(p+2)$-dimensional gauged supergravity, and find the analogous
$(\mbox{asymptotically \ads{p+2}})$ solution.  The counterterms which
arise from the dual conformal theory living on the $(p+1)$-dimensional
boundary can then be applied.  Unfortunately the two-centred (and in
general $n$-centred) \adss{p+2}{D-p-2} spacetime is, as mentioned in the
introduction, a warped geometry.  Rather than being a direct product
of an asymptotically AdS space with a sphere, it takes the form of
asymptotically(\adss{p+2}{D-p-2}).  
Certain asymptotically(\adss{p+2}{D-p-2}) geometries, such as
some {\em continuous} distributions of $p$-branes, can be consistently
reduced to a solution of a $(p+2)$-dimensional supergravity
\cite{FGPW99, CGLP99}.  Indeed, the counterterm renormalisation technique was
recently applied to such a solution in \cite{BFS01}.  However, it is
not clear to us that a {\em discrete} distribution of $p$-branes, as
the $n$-centred \adss{p+2}{D-p-2} geometries describe, finds an analogue in a reduced $(p+2)$-dimensional
supergravity\footnote{This is because one would need to turn on an
{\em infinite} number of chiral fields \cite{KW99} to accommodate the candidate
solution in a reduced supergravity.}, and thus it would appear that one cannot apply the
counterterms, as originally formulated, to this class of solutions.
An alternative which is available to us is to attempt to formulate the
counterterms directly in the eleven-dimensional theory, as was
carried out in \cite{Tay01} for the Polchinski-Strassler solutions in
type IIB supergravity, and in \cite{Tay01a} for \adss{3}{3} and \adss{5}{5}.
We will not consider this here but leave it as
a further direction to pursue.

%%%%%%%%%%%%%%%%%%%%%%%%%%%%%%%%%%%%%%%%%%%%%%%%%%%%%%%%%%%%%%%%%%%%%%%%%%%%%%
%%%%%%%%%%%%%%%%%%%%%%%%%%%%%%%%%%%%%%%%%%%%%%%%%%%%%%%%%%%%%%%%%%%%%%%%%%%%%%
%%%%%%%%%%%%%%%%%%%%%% SECTION 5 %%%%%%%%%%%%%%%%%%%%%%%%%%%%%%%%%%%%%%%%%%%%%
%%%%%%%%%%%%%%%%%%%%%%%%%%%%%%%%%%%%%%%%%%%%%%%%%%%%%%%%%%%%%%%%%%%%%%%%%%%%%%
%%%%%%%%%%%%%%%%%%%%%%%%%%%%%%%%%%%%%%%%%%%%%%%%%%%%%%%%%%%%%%%%%%%%%%%%%%%%%%
%\newpage
\sect{Other Brane Throat Instantons?}
\label{other}

%%%%%%%%%%%%%%%%%%%%%%%%%%%%%%%%%%%%%%%%%%%%%%%%%%%%%%%%%%%%%%%%%%%%%%%%%%%%%%

In the introduction to this paper we mentioned that we expect the
result for the M5 brane to hold analogously for other supersymmetric
branes in ten- and eleven-dimensional supergravity.  Here we will
present preliminary results for the M2
and D3 branes using background subtraction, and a more concrete result for
the D1/D5 system with a discussion on counterterms and subtleties in
dealing with self-dual field strengths.  We finish by
briefly commenting on other nonconformal D branes and the NS5 brane.

%%%%%%%%%%%%%%%%%%%%%%%%%%%%%%%%%%%%%%%%%%%%%%%%%%%%%%%%%%%%%%%%%%%%%%%%%%%%%%

\subsection{The M2 and D3 Branes}

Like the M5 brane, the M2 and D3 branes are nonsingular and have
anti-de Sitter near-horizon geometries --- \adss{4}{7} and \adss{5}{5}
respectively.  However, the results achieved
using background subtraction have been less conclusive.  The M2
throat instanton action contains a divergent term of order $R^4$, while
the action of the D3 throat instanton contains a divergent term of
order $R^2$.

Despite this, we would argue that these results cannot provide the
complete picture, and that if we are able to apply a suitable
counterterm technique as mentioned above, we should recover results
analogous to that for the M5 brane.  To motivate this, let us analyse
whether the boundary geometries match.

The boundary metric of the two-centred \adss{4}{7} geometry
corresponding to the two-centred M2 brane is, in the limit
$R\rightarrow\infty$,
\be
ds^2 = H^{-2/3} ds^2(\bbe{3}) + H^{1/3}R^2 d\O_7^2,
\ee
where 
\ben
H=\frac{b_1^6}{R_1^6} + \frac{b_2^6}{R_2^6} .
\een
The $S^7$ radius can easily be shown to match to the one-centred
value, but the scale factor for the worldvolume coordinates is
\be
H^{-2/3} = (b_1^6+b_2^6)^{-4} [R^4 + 4\Delta_6\, aR^3\cos\t + O(R^2)],
\label{wv-factors-M2}
\ee
where $\Delta_6$ is defined as in \eqn{def-delta}.

For the two-centred \adss{5}{5} geometry corresponding to the
two-centred D3 brane, we have
\be
ds^2 = H^{-1/2} ds^2(\bbe{4}) + H^{1/2}R^2 d\O_5^2,
\ee
where
\ben
H = \frac{b_1^4}{R_1^4} + \frac{b_2^4}{R_2^4}.
\een
The worldvolume scale factor is then
\be
H^{-1/2} = (b_1^4+b_2^4)^{-1/2}[R^2 + 2\Delta_4\, aR\cos\t + O(1)],
\label{wv-factors-D3}
\ee
with $\Delta_4$ defined as in \eqn{def-delta}.

A pattern should now emerge: if we consider \eqn{wv-factors-M5},
\eqn{wv-factors-M2} and \eqn{wv-factors-D3} (\eqn{wv-factors-brill} is
an exception), we see that the higher the dimension of the sphere, the higher the power of $R$ at which the boundary geometries
do not match\footnote{The following is meant for arbitrary values of
$\Delta_k$; choosing equal charge distribution $\Delta_k=0$ results in
reducing the power of $R$ by one, hence resulting in exact matching at the
boundary for the \adss{7}{4} geometry with two equal
centres.} --- $R^0$ for the four-sphere, $R$ for the five-sphere, and $R^3$ for the
seven-sphere.  This corresponds to one power of $R$ less than the
highest order divergent term left in
each of the on-shell instanton actions after background subtraction!
We would therefore argue that one should not trust the results obtained
from background subtraction except for the M5 brane case with two
equal centres (and, of course, the Brill instanton), and that one
should expect an appropriate counterterm procedure to show
that analogous instantons exist for the M2 and D3 branes, at least in
the cases with two equal centres.

%%%%%%%%%%%%%%%%%%%%%%%%%%%%%%%%%%%%%%%%%%%%%%%%%%%%%%%%%%%%%%%%%%%%%%%%%%%%%%

\subsection{The D1/D5 System and Self-Dual Field Strengths}

The D1/D5 system of type IIB supergravity is yet another with anti-de
Sitter near-horizon geometry, in this case, \adss{3}{3}$\,\times
\, \bbe{4}$.  The action for this system is 
\be
\kappa_{10}^2 I = - \int_{ M} d^{10}x \sqrt{g} \left( R - \frac{1}{2\cdot 3!}
F_{\it 3}^2 \right) - 2\int_{\partial M} d^{9}x \, K\sqrt{h} -
\kappa_{10}^2 I_0[h]. 
\label{10d-action} 
\ee
We are interested in the solution\footnote{Since the three-form field
strength is self-dual and thus carries both ``magnetic'' and
``electric'' components, it will necessarily become complex in the
continuation to Euclidean space.  The issue of how to interpret this
remains unresolved.  This comment also applies to the self-dual
five-form field strength in the D3 brane solution above.}
\bea
ds^2 &=& H^{-1} ds^2(\bbe{2}) + H ds^2(\bbe{4}_{(1)}) +
ds^2(\bbe{4}_{(2)}),
\label{10d-metric} \\
F_{\it 3} &=& dA_{\it 2} + \star_6 \, dA_{\it 2} ,\\
A_{\it 2} &=& i H^{-1} \mbox{vol}(\bbe{2}).
\eea
The subscripts on the two $\bbe{4}$ spaces above are simply for clarity. 
Here $\bbe{2}$ is the space tangent to both branes, $\bbe{4}_{(1)}$ is
transverse to both branes, $\bbe{4}_{(2)}$ is
transverse to the D1 branes but tangent to the D5 branes, and
`$\star_6$' refers to the Hodge dual on the space
$\bbe{2}\times\bbe{4}_{(1)}$.   We will choose spherical coordinates
on $\bbe{4}_{(1)}$
\be
ds^2(\bbe{4}_{(1)}) = dr^2 + r^2 d\O_3^2,
\ee
and compactify $\bbe{4}_{(2)}$ onto a four-torus $T^4$ of volume
$V_4$.
The choice of harmonic function
\be
H= \frac{b^2}{r^2}
\ee
corresponds to the special non-dilatonic case in which the D1 and D5
brane charges are equal and proportional to $b^2$, giving the metric
\eqn{10d-metric} the geometry \adss{3}{3}$\,\times\, T^4$ with $b$ the
radius of both \ads{3} and $S^3$.  We have chosen to study this
particular case since the resulting geometry for the two-centred choice
\be
H=\frac{b_1^2}{r_1^2} + \frac{b_2^2}{r_2^2}
\ee
is a {\em direct} product of asymptotically(\adss{3}{3}) with $T^4$.
This means that we can essentially disregard the four-torus and work
in six dimensions, with a geometry analogous to the previous cases we
have studied above.  As before, $b_1$ and $b_2$ represent the charges of
the two centres, and $r_1,r_2$ are defined as in \eqn{def-r1r2}.

As we did in section \ref{RN-BH} for the Brill instanton and section
\ref{M5-brane} for the M5 throat instanton, let us now calculate the
two-centred \adss{3}{3}$\,\times\, T^4$ Euclidean action using background
subtraction.  The bulk action vanishes by virtue of the equations of
motion (for $R$) and self-duality of $F_{\it 3}$.\footnote{In the
Euclidean continuation, the standard self-duality formula becomes
$F_{\it 3}=i*F_{\it 3}$, hence leading to the problem outlined in the
previous footnote.}  Since
\ben
K\sqrt{h} = \left(\frac{1}{2}r^3H^{-1}\partial_r H + 3r^2\right)\T_3
\een
where $\int \T_3 =\O_3$, the unit three-sphere volume, we obtain
\bean
\kappa_{10}^2 I &=& -2\int_{r=R} d^9 x \, K\sqrt{h} -\kappa_{10}^2
I_0[h] \\ 
&=& - V_2 V_4 \O_3 (4R^2 + \Delta_2^2\, a^2 ) + O(R^{-1}) - \kappa_{10}^2 I_0[h],
\eean
where $\Delta_2$ is defined as in \eqn{def-delta}
and the coordinates on $\bbe{2}$ have been compactified onto a
two-torus $T^2$ of volume $V_2$, which will be taken to infinity after
background subtraction.
Choosing the one-centred action 
\be
\kappa_{10}^2 I_0(R) = -4 V_2 V_4 \O_3 R^2 
\label{ads-3-3-one}
\ee
as the ``zero-point'', we obtain the background-subtracted action
\ben
\kappa_{10}^2 I(R) = -V_2 V_4 \O_3 \Delta_2^2\, a^2 + O(R^{-1}) .
\een
Similarly to the M5 brane throat, the action only vanishes in the
limit $R\rightarrow\infty$ 
%%%% Correction
followed by $V_2,V_4\rightarrow\infty$ if $\Delta_2$ vanishes, that is, in the
case of equal charges.  So, using background subtraction, there exists
a zero-temperature instanton describing the splitting of one D1/D5
throat into two equally charged D1/D5 throats.  Note that the background
subtraction result for splitting into two throats of unequal charge
distribution ($\Delta_2\neq 0$) again does not contradict the result of
\cite{MMS98}, nor the more detailed analysis of \cite{SW99}.

How trustworthy is background subtraction in this case?  The
worldvolume coordinate scale factor for the two-centred geometry is
\be
H^{-1} = (b_1^2+b_2^2)^{-1}[R^2 + 2\Delta_2\, aR\cos\t +
(1-4\cos^2\t(1-\Delta_2^2))a^2] + O(R^{-1}),
\ee
and does not match the one-centred value $R^2/b^2$  for any value of
$\Delta_2$.  However, we would argue, as for the Brill instanton
\eqn{wv-factors-brill}, that the result obtained by background
subtraction for the equal-centred case is correct. Perhaps it
is the low dimension of the transverse sphere, and hence the low exponent
of the highest divergent term in the action, which results in the
correct result via background subtraction.  

As expounded in the previous section, the only rigorous way of
calculating the action is to use counterterm subtraction, rather than
background subtraction.  \ads{3} corresponds to $p=1$ in the action
\eqn{ct-action}, giving the only relevant counterterms
\be
I_{ct}(\mbox{\ads{3}}) = -\frac{2}{\kappa_3^2}\int_{\partial AdS_3} d^2 x \,
\frac{\sqrt{h}}{b} + I_{log} .
\ee
For the one-centred \adss{3}{3}$\,\times\, T^4$ spacetime, $I_{log}$ again
vanishes and, using that $\kappa_{10}^2 = \kappa_3^2\, V_2V_4b^2\O_3$, we have
\be
I_{ct}(\mbox{\adss{3}{3}}\times T^4) = -\frac{V_2V_4\O_3}{\kappa_{10}^2} 2R^2.
\label{ct-action-D1/D5}
\ee
This is half the value of the one-centred \adss{3}{3}$\,\times\, T^4$
action \eqn{ads-3-3-one} calculated above!  How do we account for this
discrepancy?  

In general, varying a given action involves integrating by parts to
obtain a bulk term and a boundary term.  Forcing the bulk term to
vanish determines the equations of motion for the system, and the
vanishing of the boundary term determines the boundary conditions
which the fields must satisfy for that particular action.

However, it can happen that there is a solution which obeys these
equations of motion derived from the action but which does not obey
the specified boundary conditions.  In these cases, to compute the
correct action one needs to add boundary terms to the original action
such that the variation of this modified action still reproduces the
same equations of motion, but now determines boundary conditions which
coincide with the given solution.  

In our particular case we should consider the variation of the $F^2$
term in the bulk action:
\be
\delta \, \int_M F \wedge *F = 2\int_M d(\delta A) \wedge *F = 
\pm 2\int_M \delta A \wedge d*F + 2\int_{\partial M} \delta A \wedge *F .
\ee 
The choice of sign in the bulk term above corresponds to the rank of
$A$.  If the boundary term does not vanish when evaluated on a
particular solution, then this suggests that to obtain the correct
value of the on-shell action, one should really add to the action the
boundary term
\be
-2\int_{\partial M} A \wedge *F .
\ee
Of course in the case of \adss{3}{3} there are some subtleties due to
the lack of a covariant action incorporating the self-dual $F_{\it
3}$, however, it turns out that
\be
-\frac{1}{2} \left( -2\int_{\partial M} A_{\it 2} \wedge *F_{\it 3}\right) = 2  V_2 V_4 \O_3 R^2
\ee
is precisely the right term to add to $I_0$ \eqn{ads-3-3-one} in order
to equate it with $I_{ct}$
\eqn{ct-action-D1/D5}.\footnote{Although
it is not clear to us why this should be so, taking $F_{\it 3}^2 = |
dA_{\it 2}^2 | + | \star_6 \, dA_{\it 2}^2 |$ rather than zero also
gives precisely the right contribution to agree with the counterterm
result. }

As for counterterms in the two-centred case, 
recent progress in \cite{Tay01a} seems to indicate that the
counterterm renormalised actions for both the two-centred \adss{3}{3}
and \adss{5}{5} geometries do indeed vanish.

%%%%%%%%%%%%%%%%%%%%%%%%%%%%%%%%%%%%%%%%%%%%%%%%%%%%%%%%%%%%%%%%%%%%%%%%%%%%%%

\subsection{Nonconformal D Branes and the NS5 Brane}

Although the near-horizon geometry of the supersymmetric D$p$ brane ($p\neq~3$)
solutions of type II supergravity is only conformal to
\adss{p+2}{8-p} (or $\bbe{(6,1)}\times S^3$ for $p=5$), one can still
apply the same analysis above by working in what is called the {\em
dual} frame \cite{BST98}.  This is defined as the frame in which the metric and
the dual $(8-p)$-form field strength couple to the dilaton in the same
way, and it is in this frame that all D$p$ branes (except for $p=5$)
have \adss{p+2}{8-p} as their near-horizon geometry.  It has also been
argued \cite{BST98} that the dual frame is holographic in the
sense that taking the decoupling limit of the D$p$ brane solution
leads directly to a supergravity description, so it would be
convenient to compute quantities in this frame in order to easily
make statements about the corresponding conformal field theory.  

One final brane for which it might be interesting to perform the same
analysis is the supersymmetric NS5 brane in type IIA/B supergravity.
Its near-horizon geometry is no longer anti-de Sitter, rather it is
$S^3\times\bbr{}\times\bbr{5,1}$, and the string theory in the NS5
brane background is conjectured \cite{ABKS98} to be dual to the NS5
brane worldvolume theory.  This worldvolume theory, often referred to
as a ``little string theory'', turns out to be in an appropriate limit
a six-dimensional
superconformal field theory with $\caln$=(2,0) for type
IIA,\footnote{This is the same $(2,0)$ theory dual to M theory on
\adss{7}{4}, see \cite{AGMOO99} and
references therein for further details.  We thank T.\ Dasgupta for
clarifying this point.}  and $\caln$=(1,1) for type
IIB.  One can as before construct supersymmetric multi-centred
near-horizon geometries, and compute the appropriate Euclidean action
evaluated on these solutions.  

We expect that similar results will follow for the branes discussed in
this subsection, but leave this for future work.  

%%%%%%%%%%%%%%%%%%%%%%%%%%%%%%%%%%%%%%%%%%%%%%%%%%%%%%%%%%%%%%%%%%%%%%%%%%%%%%
%%%%%%%%%%%%%%%%%%%%%%%%%%%%%%%%%%%%%%%%%%%%%%%%%%%%%%%%%%%%%%%%%%%%%%%%%%%%%%
%%%%%%%%%%%%%%%%%%%%%% DISCUSSION %%%%%%%%%%%%%%%%%%%%%%%%%%%%%%%%%%%%%%%%%%%%
%%%%%%%%%%%%%%%%%%%%%%%%%%%%%%%%%%%%%%%%%%%%%%%%%%%%%%%%%%%%%%%%%%%%%%%%%%%%%%
%%%%%%%%%%%%%%%%%%%%%%%%%%%%%%%%%%%%%%%%%%%%%%%%%%%%%%%%%%%%%%%%%%%%%%%%%%%%%%
%\newpage
\sect{Speculations}
\label{discussion}

%%%%%%%%%%%%%%%%%%%%%%%%%%%%%%%%%%%%%%%%%%%%%%%%%%%%%%%%%%%%%%%%%%%%%%%%%%%%%%

In his original paper \cite{Bri92}, Brill conjectured the existence of
an instanton which describes the fragmentation of the complete extremal
\rn wormhole and agrees with the \adss{2}{2} instanton in the interior
throat region.   Unfortunately this has not yet been realised.
Nevertheless, we would similarly conjecture that an instanton which
describes the splitting of the M5 brane with two equal centres, and
not just its two-centred \adss{7}{4} throat, also exists.

If this is true, then drawing from the analogy we made in the
introduction, an interpretation would be that the exact eigenstate
of the full M theory ``Hamiltonian'', which we
label naively as corresponding to the M5 brane, is in fact a coherent
superposition of the quantum states $\ket{\psi_n}$ associated to each of
the $n$-centred M5 branes which reside in the same charge sector of the
supersymmetry multiplet, so that charge conservation holds. 
Intuitively, one can motivate this since the BPS property of the branes means
that there are no forces involved in separating them, and thus the
two-centred or in general $n$-centred geometries with $n$ separate
stacks of branes are equally stable configurations with the same total
energy.  
The question which remains from our analysis is whether the ``M5 brane''
eigenstate also superposes states which correspond to a {\it
non-uniform}, or unequal, distribution of branes in the stacks, as in
the case of the extremal \rn black hole.  

What does all this imply for the dual conformal field theory?  The
AdS/CFT correspondence tells us that M theory on \adss{7}{4} is
dual to a six-dimensional $\caln$=(2,0) superconformal field
theory \cite{Mal97}, about which little is known.  Since, by the
results of section \ref{other}, we expect a similar interpretation to apply to
the \adss{5}{5} throat of the D3 brane, we will discuss, in the
following paragraph, the correspondence in terms of \adss{5}{5} and
$\caln$=4 super Yang-Mills, which is much better understood.  Nevertheless, one
should read the comments below as applying to any pair of anti-de
Sitter space and conformal field theory related by the correspondence,
in particular, to \adss{7}{4} and the $(2,0)$ superconformal field theory.

Now string theory on \adss{5}{5} with one centre is dual to $\caln$=4 Yang-Mills at the superconformal point, where the vacuum
expectation values of all the scalar fields 
vanish.  Similarly, string theory on \adss{5}{5} with more than one
centre (for clarity we will discuss the case of two centres) is dual to the
conformal field theory at a point in its moduli space where some of
the scalar fields have acquired a nonzero vacuum
expectation value, hence breaking the gauge group in a manner
described in \cite{KW99}.  That the metric on two-centred \adss{5}{5}
interpolates between one-centred \adss{5}{5} at $r\approx\infty$ and a
two-centred geometry at $r\approx 0$, corresponds to the
renormalisation group (RG) flow from $\caln$=4 Yang-Mills at the
superconformal fixed point in the ultraviolet to $\caln$=4 Yang-Mills
at some fixed point in the infrared.\footnote{For one-centred \adss{5}{5} the
flow is trivial.}  

However, this is not what we believe the {\em instanton}
on the supergravity side implies for the conformal field theory.  As
touched on in the introduction, the
fact that the transition amplitude between the one-centred and
two-centred \adss{5}{5} spaces is nonvanishing implies that the true
vacuum of $\caln$=4 Yang-Mills is actually a superposition of
vacua in which the scalar fields have different
vacuum expectation values.  This is depicted in the figure below.  On
the other hand, one expects the vacua above to be {\em exact} vacua of
the full quantum $\caln$=4 Yang-Mills theory, and hence their
superposition to be forbidden by the cluster decomposition principle.
How can we understand this contradiction?

\begin{figure}[h]
\begin{center}
\epsfig{file=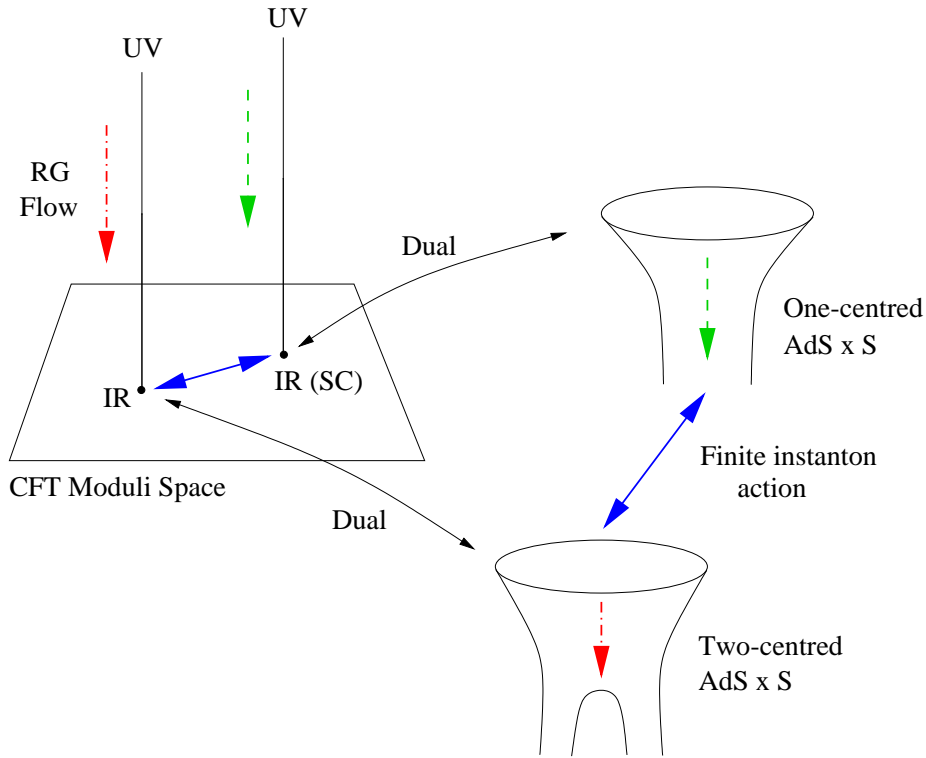, height=10cm} 
\caption{On the left hand side is the CFT picture, with the
infrared fixed points shown, one of which is superconformal (SC). The
dual AdS picture is shown on the right hand side.  The dashed (green)
and dashed/dotted (red) arrows represent RG flow from the ultraviolet
(UV) to the infrared (IR) for the one-centred and two-centred
\adss{}{} spaces respectively, while the plain (blue) arrow represents
the nonzero transition amplitude between the one-centred and
two-centred states in both pictures.}  
\end{center}
\end{figure}

One should keep in mind that the result presented in this paper was
calculated in semi-classical supergravity, and that while this serves as a
low-energy approximation to M theory, it may be that M theory finds some way of resolving the contradiction by suppressing
the tunneling, so that the effective contribution of such instantons
to the amplitude vanishes.  Since the full M theory is not yet known,
%%%% Correction: toned down from `we believe´!
it is possible that some as-yet-unknown symmetry not present in the low-energy
supergravity description may impose a new superselection rule,
preventing the tunneling process.  It would be very interesting to
investigate just how such a suppression might occur.  Since the
contradiction only arises in applying the AdS/CFT duality, one could
also question if there is a subtlety in its application here to
supergravity.  
However, we are inclined to believe that the correspondence holds
true, and that the tunneling will be suppressed in the full M theory.
Yet another possibility is that the tunneling is
instead suppressed even at the supergravity level by the presence of
fermionic zero modes\footnote{We thank M.\ Bianchi for pointing this
out, and for discussions regarding this.}.  This issue is presently
under investigation and we hope to report on it soon, 
for both the Brill instanton and the M5 throat instanton.  

Finally, we would like to comment on the implications of our results
for the issue of black hole entropy.  Some years ago, it was claimed
\cite{GK94,HHR94,Tei94} using semiclassical topological arguments,
that {\it all} extremal black holes, regardless of horizon area,
should have zero entropy, and that the Bekenstein-Hawking area law
applies only to nonextremal black holes.  Since the Hawking
temperature of extremal black holes vanishes, the zero entropy claim
is supported by the third law of thermodynamics in its strongest form.
While the extremal M5, M2 and D3 branes are in accordance with both
the area law and the zero entropy claim due to their vanishing horizon
areas, the extremal \rn black hole forces a contradiction. 

On the other hand, string theory appears to favour the area law, and
the thesis that extremal black holes should be obtained as limits of
their nonextremal counterparts.  The counting of the microscopic
quantum degrees of freedom corresponding to the black hole entropy was
pioneered by Strominger and Vafa in \cite{SV96}.  When applied to the
\rn black hole, the string analysis results in the expected nonzero
Bekenstein-Hawking entropy \cite{MS96}.  There was an attempt to
resolve this apparent contradiction between the macroscopic and
microscopic counting of the entropy in \cite{DDR96}. 

As mentioned above, the extremal M5, M2 and D3 branes each have
vanishing horizon area and hence vanishing entropy.  This agrees with
the fact that each of the above branes is BPS (supersymmetric), and
thus there should be no quantum corrections to the semiclassical
entropy as given by the Bekenstein-Hawking area law.  Furthermore, an
object with vanishing entropy should truly be {\it fundamental}, in
the sense that it has no constituent degrees of freedom.  The extremal
M5, M2 and D3 branes can indeed be considered to fall into this
category. 

%%%%%%%%%%%%%%%%%%%%%%%%%%%%%%%%%%%%%%%%%%%%%%%%%%%%%%%%%%%%%%%%%%%%%%%%%%%%%%
%%%%%%%%%%%%%%%%%%%%%%%%%%%%%%%%%%%%%%%%%%%%%%%%%%%%%%%%%%%%%%%%%%%%%%%%%%%%%%
%%%%%%%%%%%%%%%%%%%%%% ACKNOWLEDGMENTS %%%%%%%%%%%%%%%%%%%%%%%%%%%%%%%%%%%%%%%
%%%%%%%%%%%%%%%%%%%%%%%%%%%%%%%%%%%%%%%%%%%%%%%%%%%%%%%%%%%%%%%%%%%%%%%%%%%%%%
%%%%%%%%%%%%%%%%%%%%%%%%%%%%%%%%%%%%%%%%%%%%%%%%%%%%%%%%%%%%%%%%%%%%%%%%%%%%%%

%\newpage
\section*{Acknowledgments}

S.N. would like to thank M.~Bianchi, G.~W.~Gibbons, H.~S.~Reall,
K.~Skenderis, M.~M.~Taylor-Robinson and  P.~K.~Townsend for
interesting discussions. 
In particular, S.N. is grateful to D.~Mateos for many valuable
comments, many more enlightening conversations, and for a careful
reading of the manuscript.  
Finally, we would like to thank a referee for useful and insightful comments.
The work of S.N. was supported by Churchill College Cambridge and
an ORS Award.

%%%%%%%%%%%%%%%%%%%%%%%%%%%%%%%%%%%%%%%%%%%%%%%%%%%%%%%%%%%%%%%%%%%%%%%%%%%%%%
%%%%%%%%%%%%%%%%%%%%%%%%%%%%%%%%%%%%%%%%%%%%%%%%%%%%%%%%%%%%%%%%%%%%%%%%%%%%%%
%%%%%%%%%%%%%%%%%%%%%% BIBLIOGRAPHY %%%%%%%%%%%%%%%%%%%%%%%%%%%%%%%%%%%%%%%%%%
%%%%%%%%%%%%%%%%%%%%%%%%%%%%%%%%%%%%%%%%%%%%%%%%%%%%%%%%%%%%%%%%%%%%%%%%%%%%%%
%%%%%%%%%%%%%%%%%%%%%%%%%%%%%%%%%%%%%%%%%%%%%%%%%%%%%%%%%%%%%%%%%%%%%%%%%%%%%%
%\newpage

%%%%%%%%%%%%%%%%%%%%%%%%%%%%%%%%%%%%%%%%%%%%%%%%%%%%%%%%%%%%%%%%%%%%%%%%%

\end{document}